\documentclass[10pt]{amsart}
\usepackage[english]{babel}
\usepackage{amsmath,amsthm}
\usepackage{amsfonts}
\usepackage{graphicx}

\newcommand{\mx}{{\rm{mx}}}
\newcommand{\bx}{\mathbf{x}}
\newcommand{\bu}{\mathbf{u}}
\newcommand{\rmd}{\rm{d}}
\newcommand{\rme}{\rm{e}}

\begin{document}

\title[Interlaboratory consensus building challenge]{Interlaboratory consensus building challenge}%
\author{G Mana}%
\address{INRIM -- Istituto Nazionale di Ricerca Metrologica, Str. delle cacce 91, 10135 Torino, Italy}
\thanks{g.mana@inrim.it}%
\maketitle

\section{The challenge}
The challenge is about an interlaboratory comparison which involved eleven metrology institutes \cite{Possolo_2020}. It comprises four tasks
\begin{itemize}
  \item deriving a consensus value from these results;
  \item evaluating the associated standard uncertainty;
  \item producing a coverage interval that, with 95\% confidence, is believed to include the true value of which the consensus value is an estimate;
  \item suggesting how the measurement result from NIST may be compared with the consensus value.
\end{itemize}

\section{Input data}
The input data are the measured values of the iron-59 activity, $x_i$, which is positive by definition, and the associated uncertainties, $u_i$. No information is given about correlations, degrees of freedom of the uncertainty estimate, and the range of the possible measurand values. The da\-ta di\-stri\-butions encoding the available information, without introducing uncontrolled assumptions, are independent Gaussians, having common (positive) mean $\mu$ and, to avoid neglecting dark uncertainties, standard deviations $\sigma_i$ greater than or equal to the associated uncertainties.

\section{Proposed solution}
To explain the data, I considered the following set of random-error models. For some datum -- maybe none, maybe all -- the $\sigma_i = u_i$ identity holds; the other are affected by dark uncertainties. In the first case, $x_i \sim N(x_i|\mu,u_i)$. In the second, $x_i \sim N(x_i|\mu,\sigma_i)$, where $\sigma_i \ge u_i$. The hypothesis space contains as many (mutually exclusive) models as the 2048 subsets of the measured values, the empty set and its complement included. Each subset identifies the results whose associated uncertainty is the standard deviation of the sampling distribution.

Since any of data models are uncertain, to offer evidence that they explain the results or to disprove them, the sought solution must allow for comparisons. This desiderata requires that the marginal likelihood (also termed evidence) is independent of the chosen distribution parameters (e.g., the mean or stan\-dar\-dised mean and the standard deviation or variance). Consequently, it requires that the prior distributions of the different parameterisations are proper and comply with the change-of-variable rule.

Since testable information is not given, the Jeffreys' prior, which is proportional to the volume element of the $N(x_i|\mu,\sigma_i)$ manifold equipped with a local Kullback-Leibler metric, can do the work \cite{Harney_2016}. It is
\begin{equation}\label{sdv-pdf}
 \mu,\sigma_i \sim \pi(\mu,\sigma_i|u_i) = \frac{u_i}{V_\mu \sigma_i^2} ,
\end{equation}
where $u_i \le \sigma_i$, $0<\mu$, and $V_\mu$ is "volume" of the $\mu$ subspace. The sampling distribution of $x_i$, given the mean and $u_i$ and with the unknown $\sigma_i$ integrated out, is
\begin{equation}\label{xi-lik}
 x \sim L(x|\mu,u) = u \displaystyle \int_u^{+\infty} N(x|\mu,\sigma)/\sigma^2 \, \rmd\sigma = \frac{\left(1-\rme^{-\frac{(x-\mu)^2}{2u^2}}\right)u} {\sqrt{2\pi}(x-\mu)^2} ,
\end{equation}
where I dropped the $i$ subscript.

The data likelihood, given the model $A$, is
\begin{equation}\label{lik}
 \bx \sim Q(\bx|\mu,\bu,A) =\!\! \prod_{i\in A, j\in \bar{A}}\!\! N(x_i|\mu,u_i) L(x_j|\mu,u_j) ,
\end{equation}
where $A$ is a subset of normal data and $\bar{A}$ is its complement. The marginal likelihood and the posterior distribution of the mean are
\begin{equation}\label{evidence}
 Z(\bx|\bu,A) = \frac{1}{V_\mu}\! \int_{-\infty}^{+\infty}\! Q(\bx|\mu,\bu,A)\, \rmd\mu
\end{equation}
and
\begin{equation}\label{post}
 \mu \sim P(\mu|\bx,\bu,A) = \frac{Q(\bx|\mu,\bu,A)}{V_\mu Z(\bx|\bu,A)} ,
\end{equation}
where the $V_\mu$ support and $\mu$ value are large enough to allow extending the integration to the reals for all practical purposes.

\begin{figure}
\includegraphics[width=0.6\columnwidth]{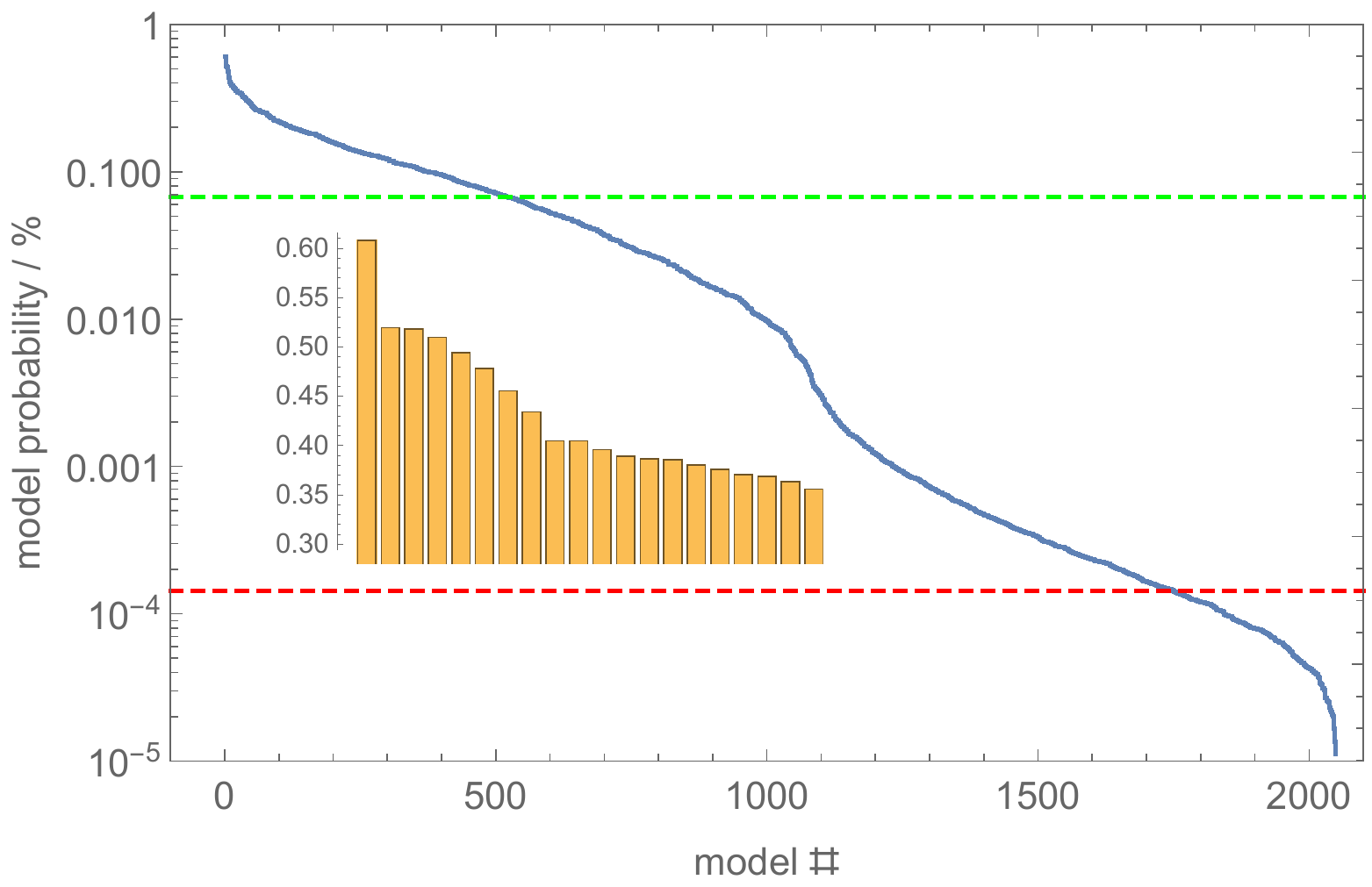}
\caption{Posterior probabilities of the subsets of normal data sorted in decreasing order. The inset shows the first 20 values. The horizontal lines are the posterior probabilities of the no-Gaussian-datum (green) and all-Gaussian-data (red) subsets.}\label{fig01}
\end{figure}

The $A_i$'s probabilities (see Fig.\ \ref{fig01}) are
\begin{equation}\label{probA}
 \rm{Prob}(A_i|\bx,\bu) = \frac{Z(\bx|\bu,A_i)}{\sum_i Z(\bx|\bu,A_i)} ,
\end{equation}
where, in the absence of additional information, I assumed equiprobable $A_i$s, which corresponds to the maximum entropy prior.

\begin{figure}
\includegraphics[width=0.6\columnwidth]{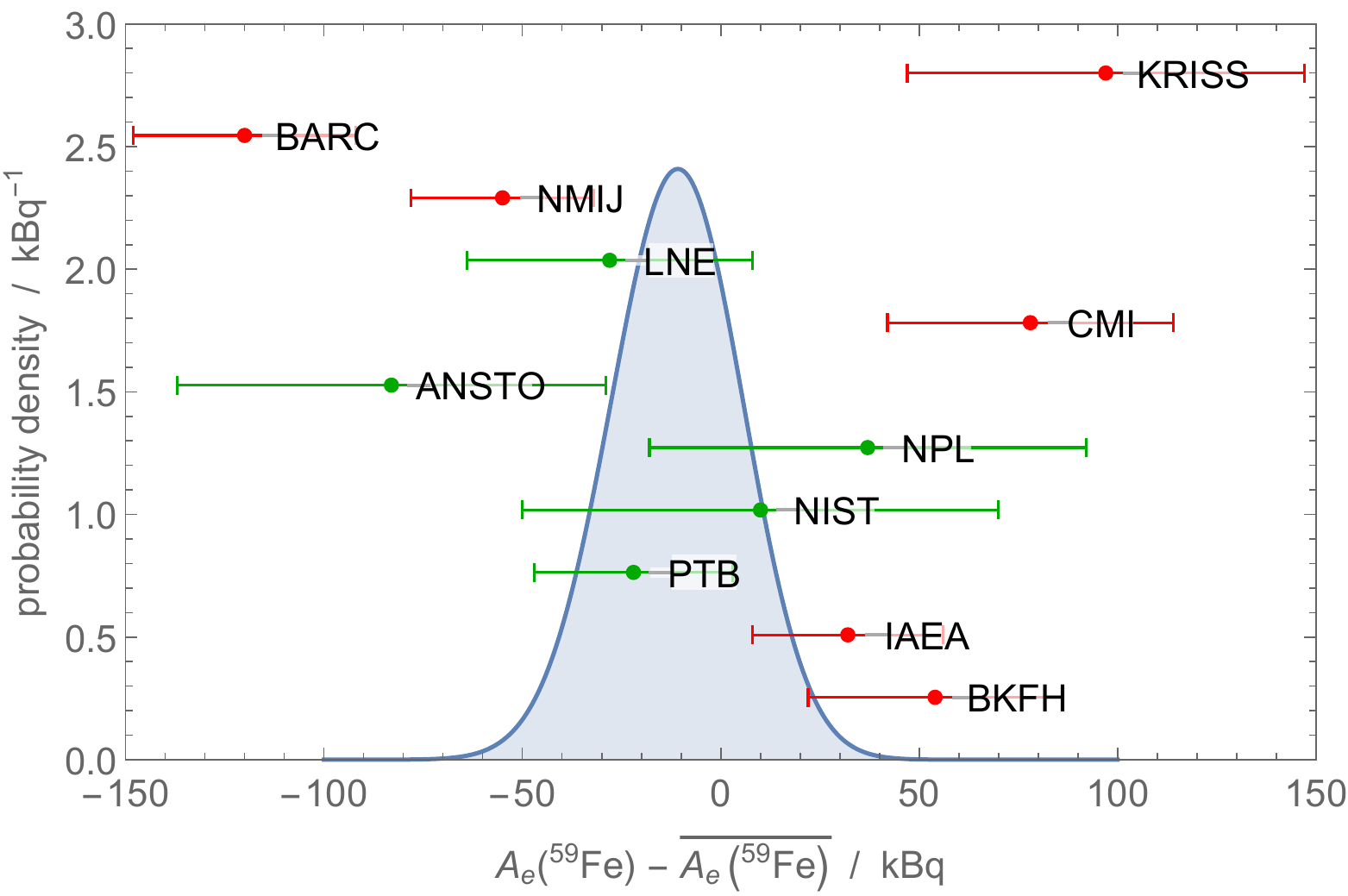}
\caption{Most probable posterior probability density for the activity of iron-59. $\overline{A_e(^{59}\rm{Fe})}=14,631$ kBq is the arithmetic mean of the data. The dots are the measured values; the lines represent the associated uncertainties. Green: Gaussian data, red: data affected by dark uncertainty.}\label{fig02}
\end{figure}

All the information about the measurand is encoded in its posterior probability density (\ref{post}) averaged over all the models,
\begin{equation}
 \mu \sim \sum_i P(\mu|\bx,\bu,A_i) \rm{Prob}(A_i|\bx,\bu) .
\end{equation}
For the sake of simplicity, I picked up the most probable model, $A_\mx$ (see Figs.\ \ref{fig01} and \ref{fig02}). Hence,
\begin{equation}\label{Pmumx}
 \mu \sim P(\mu|\bx,\bu,A_\mx)
\end{equation}
and
\begin{equation}\label{ZAmx}
 Z(\bx|\bu,A_\mx) = (61 \times 10^{-27}\, \rm{kBq}^{-10})/V_\mu .
\end{equation}
To explain the data, other models are possible. Therefore, the $A_\mx$'s evidence (\ref{ZAmx}) is a kindness to who may wish to check the $A_\mx$ explanation by the ratios of the evidence values, without having to redo the calculations. This value lets (\ref{Pmumx}) be future-proof, in that competing explanations can be compared with $A_\mx$. As an example, the evidence of the no-Gaussian-datum model is $(6.7 \times 10^{-27}\, \rm{kBq}^{-10})/V_\mu$, whereas that of the all-Gaussian-data one is $(0.01 \times 10^{-27}\, \rm{kBq}^{-10})/V_\mu$.


The choice of a consensus value is a matter of decision theory. The posterior mean, mode, and median are all equal to $14,620$ kBq. The posterior standard deviation is 16 kBq. The coverage interval that, with 95\% confidence, is believed to include the true activity is $[14,588, 14,652]$ kBq.

\begin{figure}
\includegraphics[width=0.6\columnwidth]{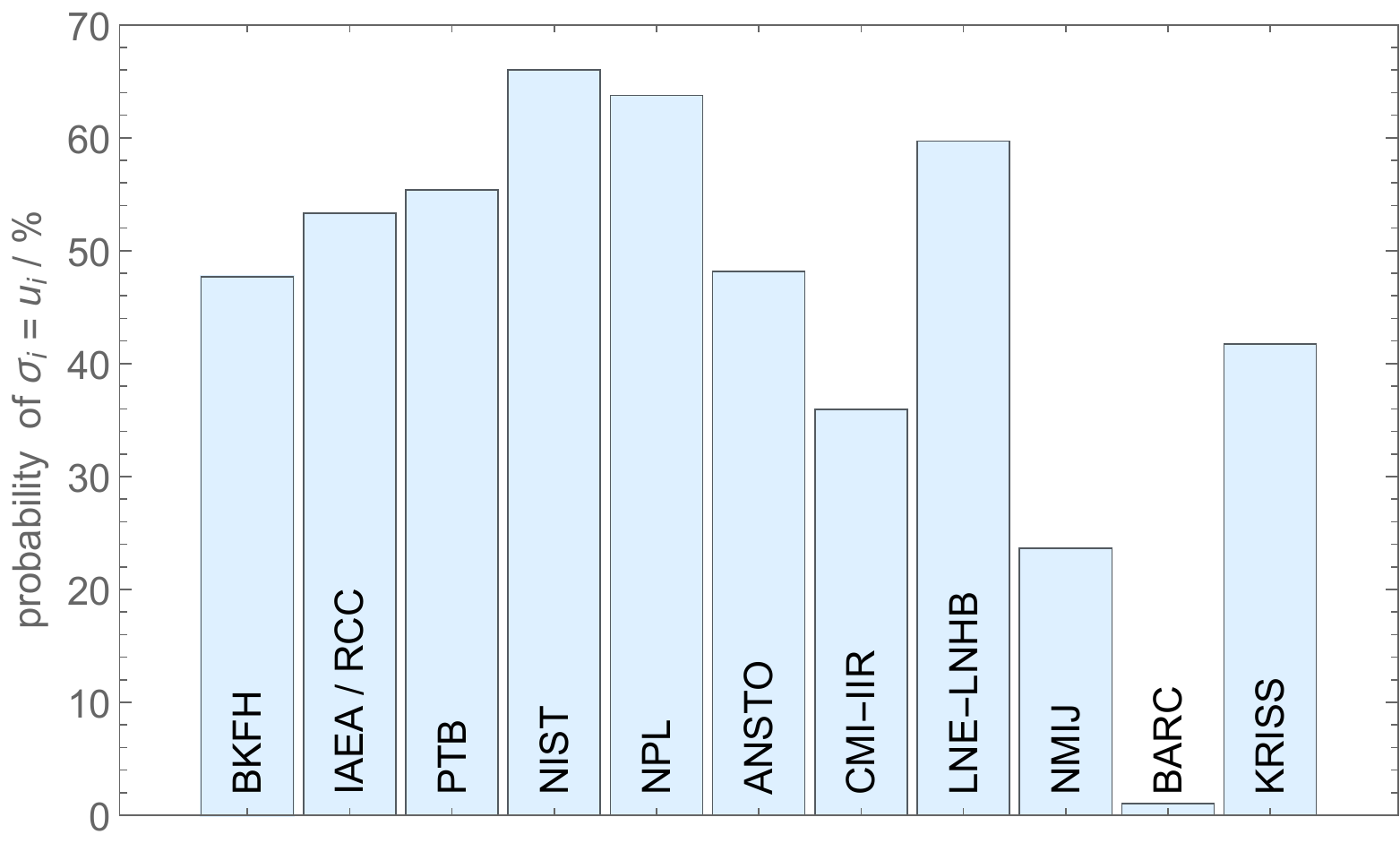}
\caption{Posterior probability that the standard deviation of the measurement result is equal to the associated uncertainty. }\label{fig03}
\end{figure}
\begin{figure}
\includegraphics[width=0.6\columnwidth]{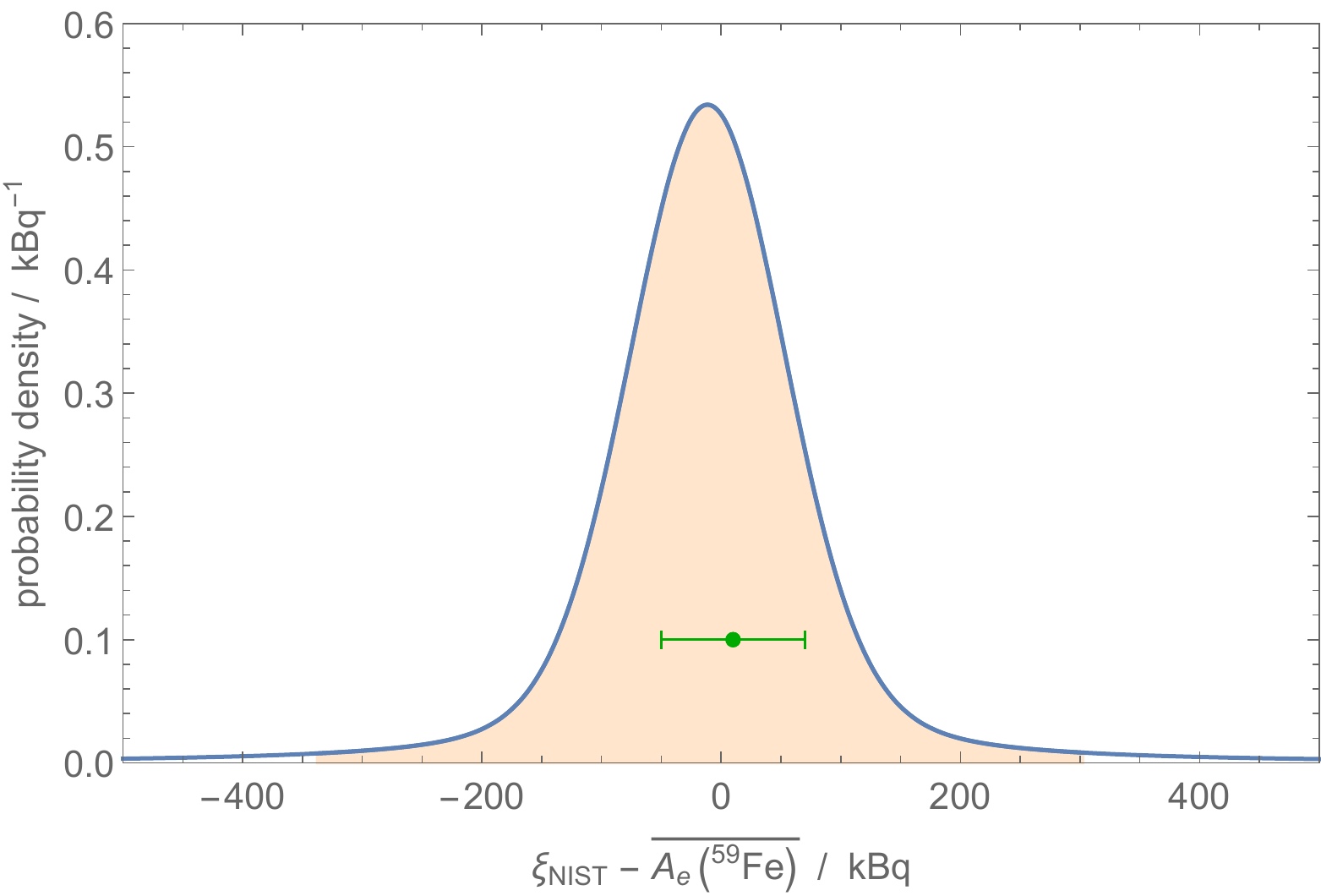}
\caption{Predictive sampling distribution of future NIST measurement results, given the data explanation $A_\mx$. The filled area is the 95\% confidence-interval. The dot is the NIST measured value; the bar is the associated uncertainty.}\label{fig04}
\end{figure}

A way to compare the measurement results with the consensus value is to evaluate the probability that its standard deviation equals the associated uncertainty. Since $A_i$ are mutually exclusive, this probability (see Fig.\ \ref{fig03}) is
\begin{equation}\label{NIST}
 {\rm Prob}(\sigma_k=u_k) = p_k = \sum_{i, x_k \in A_i} \rm{Prob}(A_i|\bx,\bu) ,
\end{equation}
For instance, the probability that the standard deviation of the NIST's measurement result is equal to the associated uncertainty is 66\%.

The $k$-th measurement result may also be compared with the consensus value (which is an estimate of the measurand value) via the predictive sam\-pling-dis\-tri\-bu\-tion (given the $A_\mx$ model, see Fig.\ \ref{fig04})
\begin{equation}\label{predictive}
 \xi|\bx,\bu,A_\mx \sim \int_{-\infty}^{+\infty} \!\!\!\big[ p_k N(\xi|\mu,u_k) + (1-p_k)L(\xi|\mu,u_k) \big] \times P(\mu|\bx,\bu,A_\mx)\, \rmd\mu ,
\end{equation}
where $\xi$ is a possible result of the laboratory $k$ measurement.

For example, the coverage interval that, with 95\% confidence, is believed to include the future NIST measurement results is from 14,293 kBq to 14,946 kBq. This interval is nearly three-times that expected from
\begin{equation}
 \xi \sim N(\xi|\mu, u_{\rm NIST}) ,
\end{equation}
that is, 653 kBq {\it vs.} 235 kBq. The difference is due to the residual 34\% probability of dark uncertainty, which is inferred from the assumed models and comparison results.


\begin{thebibliography}{1}
\expandafter\ifx\csname url\endcsname\relax
  \def\url#1{{\tt #1}}\fi
\expandafter\ifx\csname urlprefix\endcsname\relax\def\urlprefix{URL }\fi
\providecommand{\eprint}[2][]{\url{#2}}

\bibitem{Possolo_2020}
Possolo A 2020 {\em Anal. Bioanal. Chem.\/} {\bf 412} 3955--3956

\bibitem{Harney_2016}
Harney H 2016 {\em Bayesian Inference: Data Evaluation and Decisions\/}
  (Springer International Publishing)

\end{thebibliography}
\providecommand{\newblock}{}

\end{document}